\documentclass{appolb}
\usepackage{amsmath,amssymb}
\usepackage[dvipdfmx]{graphicx}

\begin{document}
\title{Time-dependent density functional studies of nuclear quantum dynamics
 in large amplitudes%
\thanks{Presented at XXII Nuclear Physics Workshop ``Marie \& Pierre Curie''}%
}
\author{Kai Wen$^1$, Kouhei Washiyama$^1$, Ni Fang$^1$, and Takashi Nakatsukasa$^{1,2}$
\address{$^1$Center for Computational Sciences and
Faculty of Pure and Applied Sciences,
University of Tsukuba, Tsukuba 305-8577, Japan\\
$^2$RIKEN Nishina Center, Wako 351-0198, Japan
} }
\maketitle
\begin{abstract}
The time-dependent density functional theory (TDDFT) provides a unified
description of the structure and reaction.
The linear approximation leads to the random-phase approximation (RPA)
which is capable of describing a variety of collective motion in
a harmonic regime.
Beyond the linear regime, we present applications of the 
TDDFT to nuclear fusion and fission reaction. 
In particular, the extraction of the internuclear potential and
the inertial mass parameter is performed using two different methods.
A fusion hindrance mechanism for heavy systems is investigated from
the microscopic point of view.
The canonical collective variables are determined by
the adiabatic self-consistent collective coordinate method.
Preliminary results of the spontaneous fission path, the potential,
and the collective mass parameter are shown for
$^8$Be$\rightarrow \alpha+\alpha$.
\end{abstract}
  
\section{Introduction}

Microscopic analysis of nuclear collective dynamics
has been of significant interest for many years.
Recently, thanks to great theoretical and computational advances,
the time-dependent density-functional calculations of
nuclear collision dynamics for heavy systems become feasible.
In this article,
we use the terminology ``time-dependent density functional theory''
(TDDFT) instead of ``time-dependent Hartree-Fock (TDHF) theory''.
We present a recent result of the real-time calculation for the heavy
systems, which suggests the fusion hindrance phenomena.
The fusion hindrance is often interpreted by the ``extra-push'' energy
which is required by a strong dissipative dynamics inside the
Coulomb barrier \cite{Swi81}.
In order to understand its complicated many-body dynamics,
it is useful to derive the internuclear
potential, the inertial mass parameter, and the energy dissipation
from the microscopic dynamics.
For this purpose, the density-constrained TDDFT calculations,
which was proposed earlier \cite{cusson85},
have been extensively performed recently \cite{UO07,UO08}.
Another method based on the classical equation of motion
was proposed in Refs. \cite{washiyama08,washiyama09}.
Using the latter method, we discuss
a possible microscopic mechanism of the fusion hindrance
in Sec.~\ref{sec:fusion_hindrance}.

Among those transport coefficients,
we put more focus on the derivation of the mass parameter in this article.
The inertial mass of nuclear collective motion has been a long-standing
problem in the nuclear structure physics \cite{BM75,RS80}.
Apparently, it is also very important for nuclear reaction dynamics.
Especially, after two nuclei touch,
the derivation of the mass is a highly non-trivial matter.
This requires extraction of the proper collective coordinates
and its conjugate momenta.
The adiabatic self-consistent collective coordinate (ASCC)
method \cite{MNM00} is capable of determining
these canonical variables by solving the self-consistent equations.
Then, the potential and the mass parameter can be uniquely determined in
the unambiguous manner.
We show out recent result for the spontaneous fission of $^8$Be,
though they are somewhat preliminary yet.

\section{Theoretical methods}

To calculate the inertial mass and the potential for the fusion reaction,
we resort herewith to two different methods:
Dissipative-dynamics TDDFT and ASCC method.

\subsection{Dissipative-dynamics TDDFT (DD-TDDFT)}
\label{sec:DD-TDDFT}

This is based on the real-time TDDFT calculation and
a mapping to the classical equation of motion.
First, we separate the whole space into that of the ``left'' and ``right''
(projectile and target).
The center of mass of the left (right) $R_L$ ($R_R$) and
the momentum of the left (right) nucleus $P_L$ ($P_R$)
is computed from the TDDFT dynamics.
Then, the left (right) mass $m_L$ ($m_R$) is estimated by
$m_L=P_L/\dot{R}_L$  ($m_R=P_R/\dot{R}_R$).
Details of the method can be found in Ref. \cite{washiyama08, washiyama09,
washiyama15}.
The important point is that complex evolution of TDDFT dynamics 
is mapped to a one-dimensional classical equation of motion:
\begin{equation}
\frac{dR}{dt} = \frac{P}{\mu}, \quad
\frac{dP}{dt} = -\frac{dV}{dR} 
                -\frac{d}{dR} \left(\frac{P^2}{2\mu} \right) 
                -\gamma \frac{dR}{dt}, \label{eq:newton}
\end{equation}
where the relative distance $R$ and the relative momentum $P$ are obtained
from $R_L$, $R_R$, $P_L$, and $P_R$.
Then, the reduced mass $\mu(R)$, 
nucleus-nucleus potential $V(R)$
and friction coefficient $\gamma(R)$, can be obtained at each $R$.

In this method, we may determine three kinds of transport coefficients,
$\mu(R)$, $V(R)$, and $\gamma(R)$.
This is an advantageous feature of the method.
However, the definition of the ``left' and ``right'' becomes somewhat
ambiguous after the two nuclei touch.
Furthermore, the assumption that $R$ and $P$ are canonical conjugate to
each other becomes questionable.
Thus, it is desirable to compare the results with other calculations.

\subsection{Adiabatic self-consistent collective coordinate (ASCC) method}
\label{sec:ASCC}

This method is based on the existence of a pair of canonical variables $(q,p)$
suitable for the description of the fusion reaction.
It leads to the basic equations of the self-consistent collective coordinate
method \cite{Mar80}
which guarantee the maximal decoupling of the collision motion
described by $(q,p)$ from the other intrinsic degrees of freedom.
Expanding those basic equations with respect to $p$ up to the second order,
we obtain the equations of the
adiabatic self-consistent collective coordinate (ASCC) method
\cite{MNM00,Hin07,Hin09,Nak12}.
Neglecting the curvature terms, it reduces to the following:
\begin{eqnarray}
&&\delta\langle \Psi(q)|\hat{H}- \frac{\partial V}{\partial q} \hat{Q}(q)  |\Psi(q)\rangle = 0, \label{chf}
\end{eqnarray}
and
\begin{eqnarray}
&&\delta\langle \Psi(q)|[\hat{H},\frac{1}{i}\hat{P}(q) ]- \frac{\partial^{2} V}{\partial q^{2}} \hat{Q}(q)  |\Psi(q)\rangle = 0, \label{RPA0} \\
&&\delta\langle \Psi(q)|[\hat{H},i\hat{Q}(q)] - \frac{1}{M(q)}\hat{P}(q)   |\Psi(q)\rangle = 0, \label{RPA}
\end{eqnarray}
where the potential $V$ is defined as
\begin{eqnarray}
V(q)= \langle \Psi(q)|\hat{H} |\Psi(q)\rangle. \label{pdef}
\end{eqnarray}
$\hat{H}$ is the total Hamiltonian, $M$ the mass parameter of collective motion.
$\hat{Q}(q)$ and $\hat{P}(q)$ correspond to local generators of
the variables $p$ and $q$, respectively, at $p=0$.
The relatively simple form of Eqs. (\ref{RPA0}) and (\ref{RPA}) is
due to the adopted approximation to neglect the curvature terms.
The original form of the ASCC equations is given in Ref. \cite{MNM00}.

Eq. (\ref{chf}) is similar to a constrained Hartree-Fock problem,
however, the constraint operator $Q(q)$ depends on the coordinate $q$
and self-consistently determined with the RPA-like equations
(\ref{RPA0}) and (\ref{RPA}).
The mass $M(q)$ corresponds to the Thouless-Valatin
mass parameters at the Hartree-Fock energy minimum,
which is known to reproduce the correct total mass for the translational
motion \cite{RS80}.
Therefore, all the quantities are determined self-consistently and
there is no {\it a priori} assumption of any kind.

After calculating the potential $V(q)$ and the mass $M(q)$,
we may transform the coordinate $q$ to the variable $R$ (distance between
two nuclei).
For the symmetric central collision, we can do this using the following
operator:
\begin{eqnarray}
\hat{R}\equiv\frac{2z}{A}\left[\theta(z)-\theta(-z)\right],\label{rdef}
\end{eqnarray}
where the collision axis is the $z$ axis.
In order to transform the mass $M(R)=M(q)(dq/dR)^2$, we need the
derivative, $dq/dR=(dR/dq)^{-1}$,
 which is obtained from the generator $\hat{P}$ in
Eq. (\ref{RPA}).
\begin{equation}
\frac{dR}{dq} =\frac{\partial}{\partial q}\langle \Psi |\hat{R}| \Psi \rangle =\langle \Psi |[\hat{R},\frac{1}{i}\hat{P}]| \Psi \rangle .
\label{mass3}
\end{equation}

In this method, the canonical variables $(q,p)$ are automatically
determined by the ASCC equations and the weak canonicity condition,
\begin{equation}
 \langle\Psi(q)|[\hat{P}(q),\hat{Q}(q)]|\Psi(q)\rangle=-i.
\label{weak}
\end{equation}
Here, $R$ is simply a measure of the collective variable $q$.
Thus, the inertial mass $M(R)$
with the coordinate $R$ describes the identical dynamics
as $M(q)$ does with the coordinate $q$.
Note that this property is not guaranteed if we assume the collective
coordinate $R$ in the beginning to define the mass $M(R)$.

\section{Application of DD-TDDFT: Fusion hindrance in heavy systems}
\label{sec:fusion_hindrance}

The synthesis of superheavy elements by heavy-ion fusion reactions
is a challenging task because of its extremely low probability.
In such fusion reactions with heavy nuclei, 
whose charge product $Z_PZ_T$ is larger than $1600-1800$, 
it has been observed that the formation of 
a compound nucleus is strongly hindered at energies around the Coulomb barrier
compared with $Z_PZ_T < 1600$ systems \cite{sahm85}.
The most probable reason behind this fusion hindrance is the occurrence of 
the quasi-fission, which involves re-separation without the formation of 
a compound nucleus after colliding nuclei touch each other.
A macroscopic fluctuation--dissipation model using a Langevin equation
has been developed \cite{ZG05,aritomo12} to analyze the competition between 
the quasi-fission and compound nucleus formation, especially in the synthesis 
of superheavy elements. Recently, the quasi-fission process was 
analyzed by microscopic TDDFT \cite{simenel12,GN12,OU14}.

\begin{figure}[t]
\centering
\includegraphics[width=0.49\linewidth,clip]
{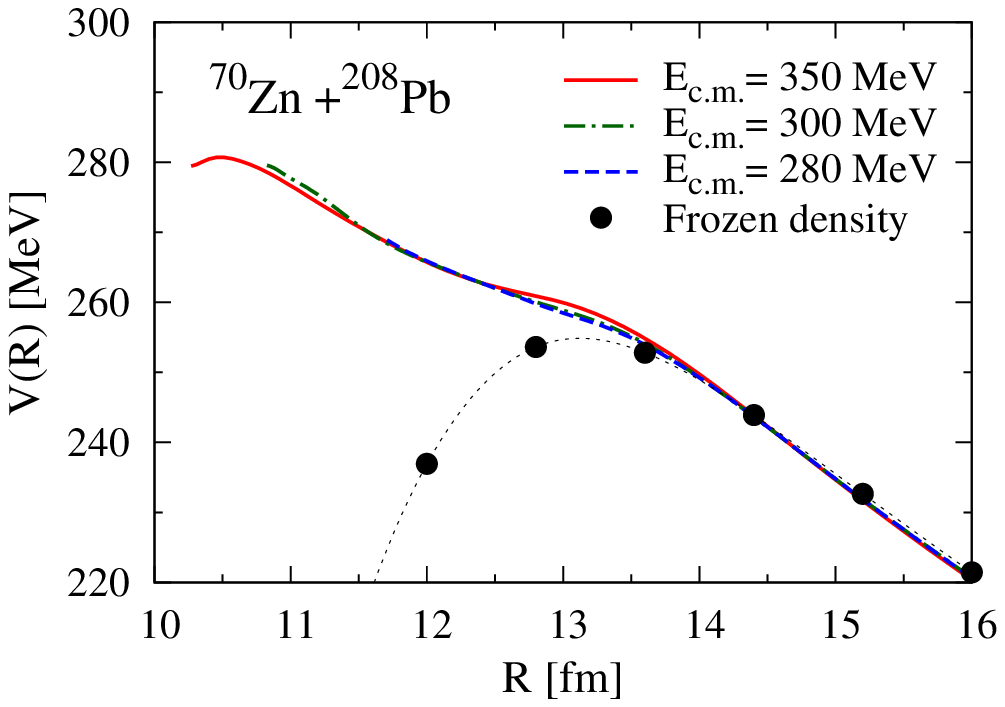}
\includegraphics[width=0.49\linewidth,clip]
{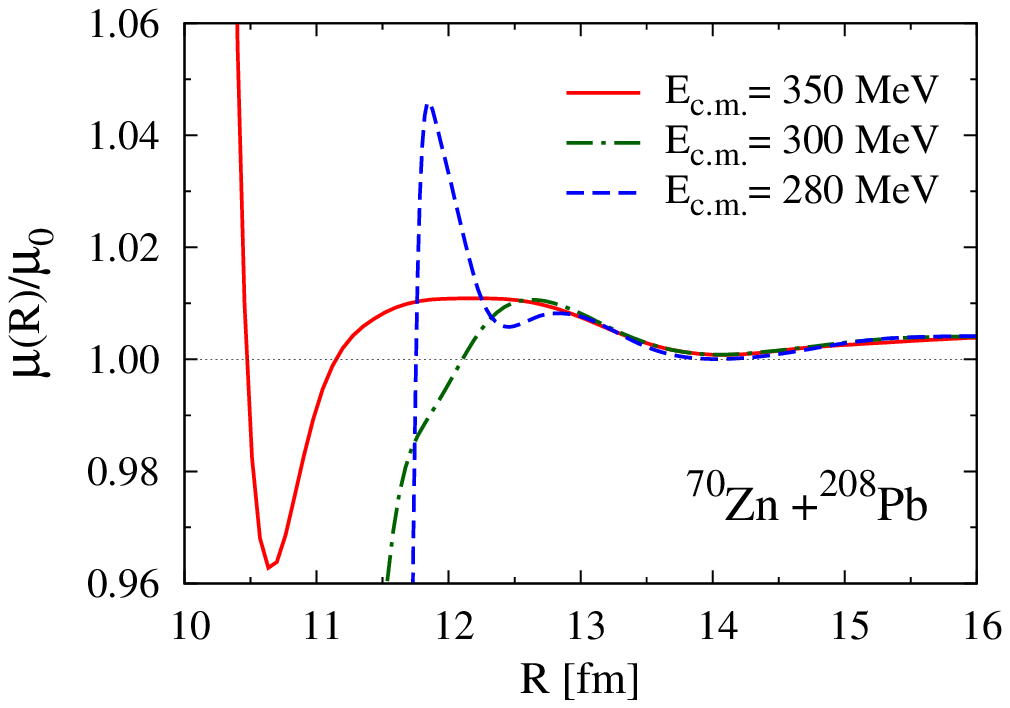}
\caption{\label{fig:DD-TDDFT} 
(Left panel)
Nucleus--nucleus potential for the $^{70}$Zn\,+\,$^{208}$Pb system. 
The lines denote the potentials extracted from the TDDFT trajectories 
at different $E_{\mathrm{c.m.}}$, 
while the filled circles obtained from the frozen density approximation.
(Right)
Calculated inertial mass $\mu(R)$ at different $E_{\mathrm{c.m.}}$.
The Skyrme SLy4d parameter set is used in the TDDFT simulation.
 }
\end{figure}

In this section, we show the microscopic analysis of the fusion hindrance
using the techniques in Sec.~\ref{sec:DD-TDDFT},
from the point 
of view of energy dissipation and dynamical nucleus--nucleus potential.
The left panel of Fig.~\ref{fig:DD-TDDFT} shows the nucleus-nucleus potential
for $^{70}$Zn\,+\,$^{208}$Pb system obtained from the TDDFT simulations at 
$E_{\mathrm{c.m.}}=350$, 300, and 280\,MeV, and the frozen density potential 
that is obtained from the energy of the total system at each $R$
while holding the projectile and target densities frozen to be 
their ground-state densities. 
We find that an ordinary barrier disappears in the extracted potentials 
at all energies, and as a result, monotonic increase of the potential is seen.
These properties are observed only in heavy systems \cite{washiyama15}.

We analyze the origin of fusion hindrance. In the present method,
the extra-push energy can be identified with the contribution of 
the increase in potential observed in Fig.~\ref{fig:DD-TDDFT} (left) and 
of the accumulated dissipation energy computed from the friction coefficient 
$\gamma(R)$.
Table~\ref{tab:origin} shows the contributions of increase in potential 
from the frozen density potential and of the accumulated dissipation energy 
to the extra-push energy for selected systems at the fusion threshold energy. 
It is clear that $\Delta V$ is larger than $E_{\mathrm{diss}}$.
The larger the charge product of the system,  
$\Delta V$ is more dominant.

\begin{table}[b]
\caption{\label{tab:origin}%
Selected results of contributions of the increase in potential $\Delta V$
and of the accumulated dissipation energy $E_{\mathrm{diss}}$ to the extra-push energy $E_{\mathrm{extra}}$.
}
\vspace{1em}
\centering
\begin{tabular}{c|ccc}
\hline
System & $E_{\mathrm{extra}}$(MeV) & $\Delta V$(MeV) & $E_{\mathrm{diss}}$(MeV)  \\
\hline
$^{100}$Mo\,+\,$^{110}$Pd & 14.7 & 6.7 & 6.3 \\
$^{96}$Zr\,+\,$^{124}$Sn & 13.9 & 8.5 & 3.9\\
$^{96}$Zr\,+\,$^{136}$Xe & 16.0 & 9.7 & 5.1 \\
\hline
\end{tabular}
\end{table}

In the right panel of Fig.~\ref{fig:DD-TDDFT},
we show the calculated inertial mass parameter $\mu(R)$.
The energy dependence is relatively small at $R\gtrsim 13$ fm.
It seems to suggest a slight increase of $\mu(R)$ as two nuclei
approach each other.
However, at $R\lesssim 13$ fm, 
$\mu(R)$ significantly oscillates and depends on
the colliding energy $E_\textrm{c.m.}$.
It is interesting to observe that, despite of this strong energy dependence
of $\mu(R)$, the calculated potential $V(R)$ is rather universal.

\section{Application of the ASCC method: Potential and inertial mass}

Because of the unique definition of the potential and mass,
it is highly desirable to calculate these in the ASCC method.
Although the numerical application of the ASCC method to the fusion
reaction requires a large computational task,
we show here our first (preliminary) result for the simplest case,
the spontaneous fission process of $^8$Be$\rightarrow \alpha+\alpha$.
It can be regarded as those for the fusion process of two $\alpha$
at very low energy.

The model space is the three-dimensional grid space inside the
sphere of the radius of 7 fm with the square mesh size of 1 fm.
The energy density functional is the simple BKN functional \cite{BKN}.
To solve Eqs. (\ref{RPA0}) and (\ref{RPA}), we use the
finite amplitude method (FAM) \cite{NIY07,AN11},
especially the matrix FAM prescription \cite{AN13}.

To fully determine the collective variables $(q,p)$, the potential $V(q)$,
and the mass parameter $M(q)$,
an iterative procedure is required
to obtain a self-consistent solution that satisfies all the equations
(\ref{chf}), (\ref{RPA0}), and (\ref{RPA}).
The initial trial wave functions are constructed with the constrained
HF minimization with respect to a given constraint operator,
such as $\hat{R}$ and the mass quadrupole operator $Q_{20}$.
Then, the iteration is performed until it reaches the convergence
to simultaneously satisfy Eqs. (\ref{chf}), (\ref{RPA0}), and (\ref{RPA}).

\begin{figure}[t]
\includegraphics[width=0.49\columnwidth]{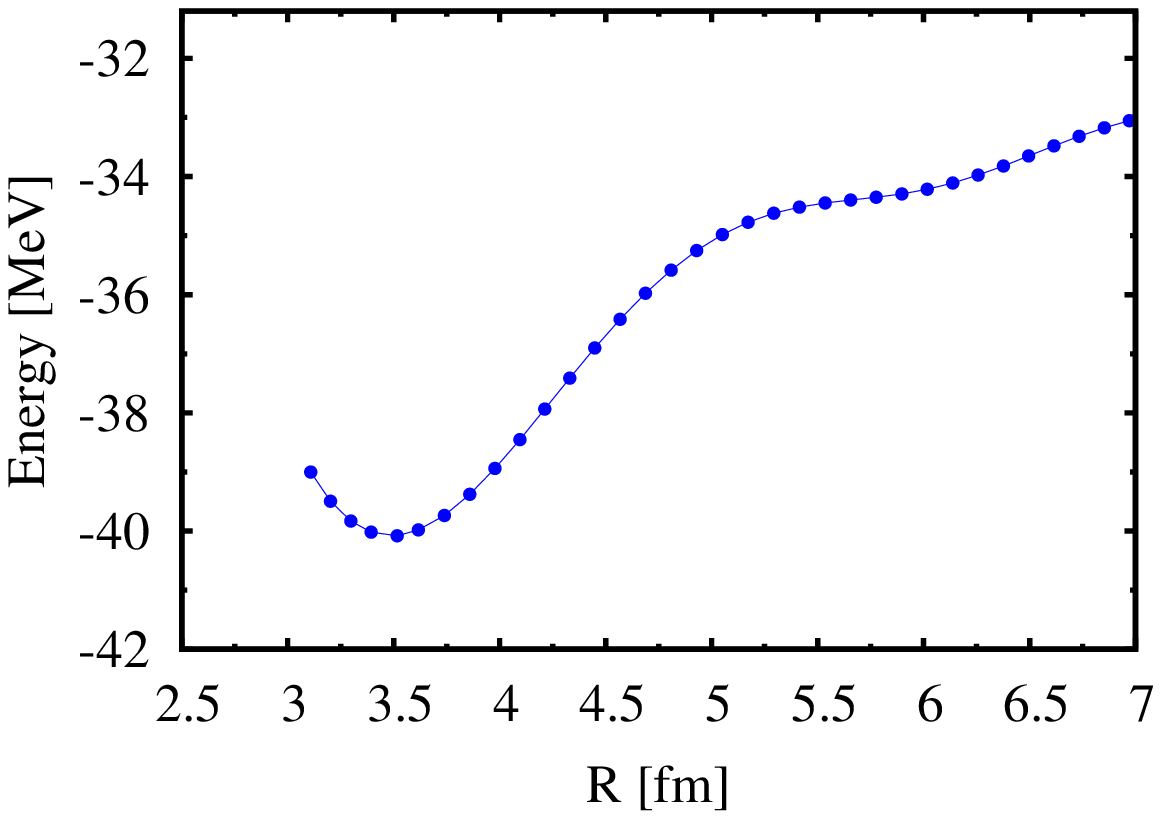}
\includegraphics[width=0.49\columnwidth]{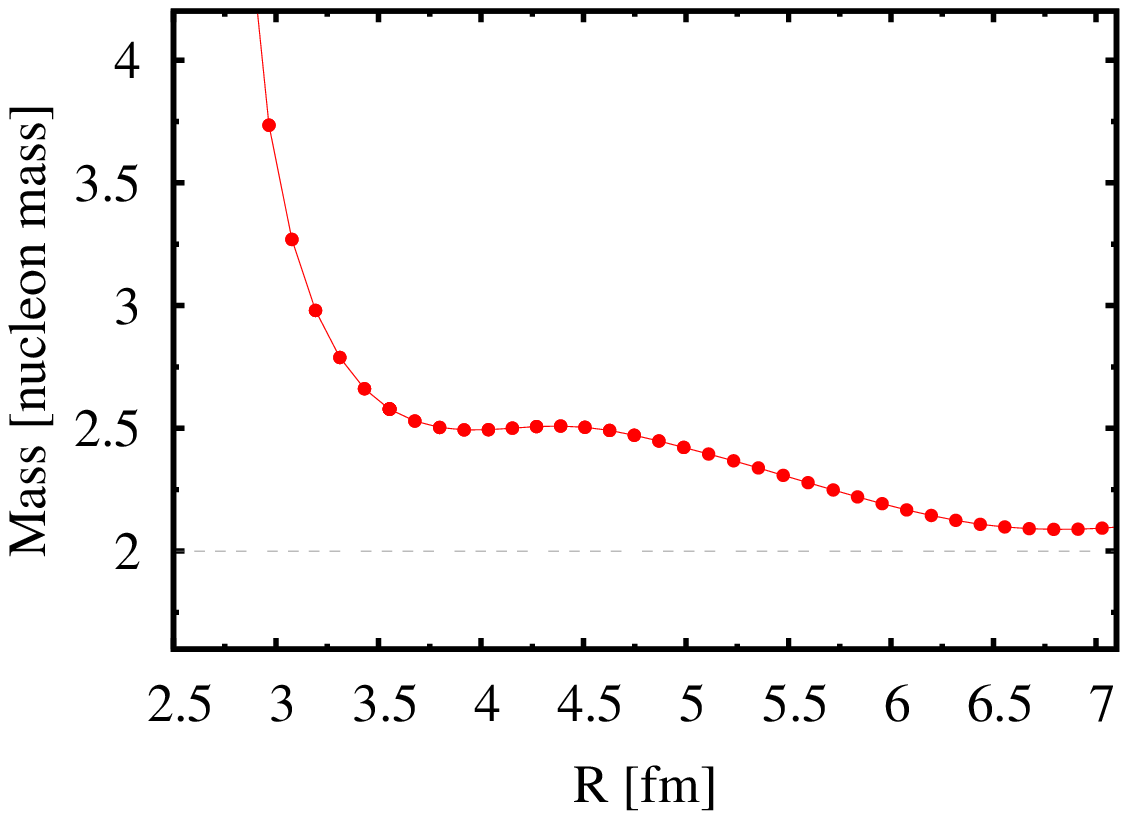}
\caption{\label{fig:ASCC}
(Left panel) Calculated potential energy as a function of the relative
distance between the two $\alpha$ particles, $V(R)$.
(Right)
Calculated inertial mass parameter
for the relative motion between two $\alpha$ particles $M(R)$
as a function of the relative distance $R$.
The vertical axis is normalized to the nucleon mass $m$.
}
\end{figure}

The obtained potential and the inertial mass are shown in
Fig.~\ref{fig:ASCC}.
The minimum around $R=3.5$ fm correspond to the HF solution of $^8$Be.
The potential shows somewhat funny behaviors at large $R$, which is
due to the restriction of the model space.
The Coulomb barrier should be around $R=6.3$ fm.
At larger $R$, we obtain the mass parameter $M(R)\approx 2 m$ which
correspond to the reduced mass of two-$\alpha$ system.
The mass $M(R)$ shows a rapid increase at $R<3.5$ fm, however,
we should not take this seriously.
In fact, we have found in this interior region, the solution of
the RPA-like equations (\ref{RPA0}) and (\ref{RPA}) corresponds to
the state in the continuum, above the bound threshold.
Therefore, the result in this region strongly depends on
the choice of the model space.
Nevertheless, even at $R>3.5$ fm,
as the two $\alpha$ particles get closer to each other,
the inertial mass $M(R)$ gradually increases.
This is similar to the behavior we have observed in the right
panel of Fig.~\ref{fig:DD-TDDFT}.
The detailed analysis is under progress.

\section{Summary}
We have discussed the applications of the TDDFT to nuclear fusion
reaction.
The fusion hindrance phenomenon is studied with the real-time
simulation and the DD-TDDFT method to extract the potential, the mass,
and the friction parameter.
This study suggests that the fusion hindrance is mainly due to the
potential increase inside the Coulomb barrier.
The mass parameter determined by this method shows a strong
energy dependence.

A proper definition of the canonical variables suitable for
the nuclear fusion  process, especially after the two nuclei touch
each other, is important and challenging.
Their self-consistent and unambiguous definition can be given by the
TDDFT dynamics itself, and
the ASCC method provides a feasible approach to this at low-energy collision.
In this method, the mass parameter and the potential are uniquely determined.
The numerical calculation has been performed for the simple case
of the fission of $^8$Be into two $\alpha$'s.
The obtained mass coincides the reduced mass of $2m$ when the two
$\alpha$ particles are far away.
As they approach to each other, the mass gradually increases.

The mass parameter in TDDFT contains effects of the time-odd mean fields
which are known to be necessary to reproduce the correct total mass
for the translational motion \cite{RS80}.
This is not achieved neither by the Inglis-Belyaev cranking formula, nor
the generator coordinate method with real coordinates \cite{RS80}.
Therefore, the mass parameters determined in the ASCC method has
a clear advantage.
The drawback is its computational task to solve the RPA-like equations
on every point on the collective path determined by Eq. (\ref{chf}).
The analysis and applications to heavier systems are under progress.

\section{Acknowledgement}
This work is supported by KAKENHI (Grants No. 2415006 and 25287065)
and by ImPACT Program of Council for Science, Technology and Innovation
(Cabinet Office, Government of Japan).
Numerical calculations were performed in part using COMA at the CCS,
University of Tsukuba, and RICC at RIKEN.

\end{document}